\documentclass[graybox]{svmult}
\usepackage{mathptmx, helvet, courier, makeidx, graphicx, multicol, type1cm}
\usepackage{keyval,cite}
\usepackage[bottom]{footmisc}

\def\be {\begin{equation}}
\def\ee {\end{equation}}
\def\bea {\begin{eqnarray}}
\def\eea {\end{eqnarray}}
\def\bc {\begin{center}}
\def\ec {\end{center}}

\title*{PeV neutrinos from local magnetars}
\author{Rajat K. Dey}
\authorrunning{R. K. Dey}
\institute{Rajat K. Dey \at Department of Physics, University of North Bengal, Siliguri, West Bengal, INDIA, \email{rkdey2007phy@rediffmail.com}}
\date{\today} 

\begin{document}

\maketitle 

\abstract{
The paper estimates the flux of PeV neutrinos from magnetar polar caps, assuming that ions/protons are injected, and accelerated in these regions and interact with the radiative background. The present study takes into account the effect of the photon splitting mechanisms that should modify the radiative background, and enhance the neutrino flux at PeV energies, with a view to explain the PeV neutrino events detected in IceCube. The results indicate that in near future, a possibility of any significant excess of neutrino events from a magnetar in Milky Way is extremely low.}

\abstract*{
We estimate the flux of PeV neutrinos from magnetar polar caps, assuming that ions/protons are injected, and accelerated in these regions and interact with the radiative background. The present study takes into account the effect of the photon splitting mechanisms that should modify the radiative background, and enhance the neutrino flux at PeV energies, with a view to explain the PeV neutrino events detected in IceCube. The results indicate that in near future, possibility of any significant excess of neutrino events from a magnetar in Milky Way is extremely low. }

\section{Introduction}
\label{sec:1}
The detection of 3 very high energy (VHE) astrophysical neutrinos with the IceCube detector has recently opened up a whole new window on the energetic Universe ~\cite{aar13}. The present work proposes a viable physical model of the interactions taking place in source regions for the prediction of astrophysical neutrinos.  

The possible origins of these PeV neutrino events have been discussed by many in recent times ~\cite{lah13}, but still remains a topic of much speculation. A fundamental question is therefore being raised: what classes of astronomical objects could accelerate hadrons to very high energies, and in which types of interactions are neutrinos then produced?

In this situation, acceleration of protons in the vicinity of the surface of relatively young local neutron stars with super strong magnetic fields ($B \sim 10^{14-17}$ G), widely known as magnetars, is in fact supposed to be a possibility. The subsequent production of photomesons by interactions with radiative background proceeds, and that has already been studied by many authors. Here, we have proposed an additional contribution to the target photon fields for photomeson production from the photon splitting mechanism. The photon splitting process is expected to modify the radiative background, and enhance the neutrino flux at PeV energies from the object. 

The phase diagram $P - B$,  corresponds a phase in which the star could pass a neutrino-loud regime ~\cite{zha03}. If the spin-down power of a local magnetar for a favorable set of $P$ and $B$ at a particular evolutionary phase is consumed to accelerate protons, then the object might emit PeV muon neutrinos ($\nu_{\mu}$) through photomeson interactions. The radiative background of the star is believed to be filled mainly with soft ultraviolet (UV)-A or B photons that are in turn produced from the effect of photon splitting mechanism on magnetar's unmodified radiative background.

\section{Photon splitting process in the magnetosphere}
\label{sec:2}
The photon splitting, a QED process that splits a high-energy photon into pair of low energy photons in presence of a pure magnetic field and/or magnetized plasma. No detailed calculation on the probabilities of these processes in the current scenario ($B > B_{\rm{cr}} = 4.41 \times 10^{13}$ G with plasma) is available except a numerical calculation used in ~\cite{chi12}. The environment affects the rate by changing photo dispersion properties of the region. 

Photon splitting may occur via various possible channels which are determined by the electric field vector ($\bf{X}$), momentum vector ($\bf{q}$) associated with a photon, and the magnetic field vector ($\bf{B}$). The state {\lq 1\rq} refers to a configuration where $\bf{X}$ is perpendicular to the plane containing $\bf{q}$ and $\bf{B}$ while {\lq 2\rq} corresponds the $\bf{X}$ being parallel to their plane. For low energy background target UV-C photons ($T_{\rm{kin.}} \cong 0.1 - 0.2$ keV $<< m_{\rm e}c^{2}$) the only physical mode, $\gamma_{1} \rightarrow \gamma_{2} + \gamma_{2}$ is responsible for splitting and, thereby soften background photon spectra in the magnetized plasma. But in the absence of plasma, the physical mode, $\gamma_{1} \rightarrow \gamma_{1} + \gamma_{2}$ is the significant one for photon splitting over the channel, $\gamma_{1} \rightarrow \gamma_{2} + \gamma_{2}$. 

\section{PeV neutrinos from magnetars: Physical model}
\label{sec:3}
In particle astrophysics, it has been presumed that the VHE protons and/or ions are injected, and accelerated in surrounding regions of cosmic accelerators. These accelerated ions then interact with the radiative background and subsequently VHE neutrinos and gamma-rays are generated via dominant photo-meson interactions,

\begin{equation}
p + \gamma \rightarrow \Delta^{+} \rightarrow \left\{\begin{array}{ll}
p + \pi^{\rm o} \rightarrow p + 2\gamma \\
n\pi^{+} \rightarrow n + e^{+} + \nu_{\rm e} + \nu_{\mu} + \bar{\nu_{\mu}}
\end{array}
\right. 
\end{equation}
										   
The final products of all neutrino flavours turn into a ratio of ${\nu}_{\rm{e}} : {\nu}_{\mu} : {\nu}_{\tau} = 1:1:1$ at earth.

In magnetar's early life, the spin-down power is consumed to accelerate protons/ions, and the magnetic field driven power supplies ambient photon targets. The kinematic threshold for photo-meson interaction process in eq.(1) is determined by the accessible photon energies in the radiative field that are mostly UV type photons. 

The neutron star (NS) remnant or merger that appears from massive binary NSs ($M \sim 2{\rm M}_{\odot}$) coalescence, may form a millisecond magnetar with thin ejecta walls across polar caps in its very early phase. As the star receives huge angular momentum from the binary it possesses a rapid rotation at the moment of its birth. These magnetars also have super-strong magnetic fields ~\cite{zra13}. Here, we suggest that in the evolutionary phase of a magnetar after NSs merger, the star may transit a state when spin-down power is comparable with its magnetic power, and spin period falls in the range $200 - 500$ ms. 

An extension of previous calculations for young pulsars to magnetars reveals that protons or heavier ions undergo acceleration in the magnetar's polar caps attaining energies close to $10^{16} - 10^{17}$ eV, provided the magnetar's magnetic moment vector $\bf{\mu}$ and $\bf{B}$ parameter satisfy the strong condition; $\bf{\mu}.\bf{B} <0$. These VHE protons will interact with soft UV-A and UV-B photons close to the magnetar's polar caps, the $\Delta$ resonance state may form satisfying the kinematic threshold condition for the process in eq.(1). The energy of the modified target photons is $2.8{kT}_{\infty}(1+z_{\rm g}) \sim 0.01$ keV, where $z_{\rm g}$ $\sim 0.4$ being the gravitational red shift. Thus the proton threshold energy $\epsilon_{\rm{p},{\rm Th}}$ for the $\Delta^{+}$ resonance state ranges $\geq 3\times10^{16}$ eV. The VHE proton flux emitted from the polar cap region would therefore be
              
\begin{equation}
\Phi_{\rm{PC}}\simeq c f_{\rm d}(1-f_{\rm d}) n_{\rm o} A_{\rm{PC}}, 
\end{equation}

where $A_{\rm{PC}}$ denotes polar cap area, and it is $\eta_{\rm A} (4 \pi R^{2})$ with $\eta_{\rm A}$ accounts the ratio of polar cap area to the magnetar surface area. Earlier calculations in ~\cite{lin05} for estimating proton/ion flux in pulsar's polar caps took the parameter $\eta_{\rm A}$ as unity. The characteristic polar cap radius can be given by, $r_{\rm{PC}}= R (\Omega {R/c)}^{1/2}$, and hence $\eta_{\rm A}$ takes the form $\Omega R/(4c)$. 

It is seen from the process in eq.(1) that the charge-changing reaction goes on just $\frac{1}{3}$-rd of the reaction time, about three high-energy neutrinos (or a pair of $\nu_{\mu},\bar{\nu_{\mu}}$) will accompany with four high-energy gamma-rays on the average when a significant number of such reactions proceed successfully. The total flux of neutrinos that is originated from the disintegration of $\Delta^{+}$ resonance state will be 
\begin{equation}  
\Phi_{\nu}(r \simeq 1.2R) =  2 c f_{\xi} A_{\rm{pc}} f_{\rm d} (1-f_{\rm d}) n_{\rm o} P_{\rm c},
\end{equation}
   
with $f_{\xi}$ is $2/3$. If now the duty cycle factor $f_{\rm{dc}}$ of the muon neutrino is taken into account, the phase averaged $\nu_{\mu}$ flux on the Earth from a magnetar at a distance $D$ is given by 
\begin{equation}
\Phi_{{\nu}_{\mu},\bar{\nu_{\mu}}} \simeq  2 c f_{\xi} f_{\zeta} \eta_{\rm A} f_{\rm{dc}} f_{\rm s} f_{\rm d} (1-f_{\rm d}) n_{\rm o}\left(\frac{R}{D}\right)^{2}P_{\rm c}
\end{equation} 

The effect of neutrino oscillations is represented by the  parameter $f_{\zeta}$ (here, it is 1/2). The factor $f_{\rm s}$ is set equal to 1 for $\nu_{\mu}$.

 We now calculate numerical values for $\nu_{\mu}$ flux using the formula in the eq.(4) for a typical galactic magnetar with $D \sim 2$ kpc, $P \sim 350$ ms, $B_{15} \sim 1.5$, $T_{0.1 \; \rm{keV}} \sim 0.0255$, and $f_{\rm{dc}} \leq 0.10$ in both the cases when $\eta_{\rm A}$ is equal to (i) 1 and (ii) $\Omega R/(4c)$. We have taken star radius equal to $R = 10$ km for the present calculation. For the purpose, we choose $Z = 1$ and $f_{\rm d} = 1/2$ here.

The corresponding $\nu_{\mu}$ flux ($E^{2}\phi_{\nu_{\mu}}$) calculated out  are $6.03 \times 10^{-10}$ in $\rm{GeV cm}^{-2}s^{-1}$ for $\eta_{\rm A} = 1$. These values are $0.0009 \times 10^{-10}$ according to the case (ii) in $\rm{GeV cm}^{-2}s^{-1}$. If we compare with IceCube estimated integral PeV neutrino flux, that is, $\sim 2.4\times 10^{-9}$ $\rm{GeV cm}^{-2}s^{-1}$, these predicted values look quite low, particularly in (ii). IceCube has measured neutrino flux as $E^{2}\phi_{\nu_{\mu}+{\tilde{\nu}_{\mu}}} \sim 3\times 10^{-8}$ GeV cm$^{-2}$ s$^{-1}$ sr$^{-1}$ corresponding to neutrino energy range $0.2 - 2$ PeV. 

\section{Conclusions}

If protons reach $10 - 100$ PeV energy scale in a magnetar then their interactions with modified UV-A/B photon targets may generate PeV neutrino events with energies between $1 - 10$ PeV as observed by the IceCube experiment. The model suggests no possible indication of any statistically significant excess from the direction of any local magnetar to be observed by IceCube in near future.

\begin{acknowledgement}
This work was supported by SERB, DST, Govt. of India through its grant no. EMR/2015/001390.
\end{acknowledgement}

\end{document}